\documentstyle[12pt]{article}
\advance \topmargin by -\headheight
\advance \topmargin by -\headsep
\evensidemargin \oddsidemargin
\title{Field Theory Models without the Cosmological Constant Problem}
\author{E.I.Guendelman \thanks{GUENDEL@BGUmail.BGU.AC.IL} and
        A.B.Kaganovich \thanks{ALEXK@BGUmail.BGU.AC.IL}}
\date {Physics Department, Ben Gurion University of the Negev,
   Beer  Sheva 84105, Israel}
\begin{document}
\maketitle
\begin{abstract}
 We study field theory models in the context of a gravitational 
theory without the 
cosmological constant problem (CCP). The theory is based on the
requirement 
that the measure of integration  in the action is not necessarily
$\sqrt{-g}$ 
but it is determined dynamically through additional degrees of freedom,
like four scalar fields $\varphi_{a}$.  We study three possibilities for 
the general structure of the
theory: (A) The total action has the form  $S=\int\Phi Ld^{4}x$ where
the measure $\Phi$ is built from the scalars $\varphi_{a}$ in such a
way that the transformation $L\rightarrow L+const$
does not effect equations of motion. 
Then an
infinite dimensional shifts group of the measure fields (SGMF)
$\varphi_{a}$ by arbitrary functions of the Lagrangian density $L$
$\varphi_{a}\rightarrow \varphi_{a}+f_{a}(L)$ is recognized as the
symmetry group of the action up to an integral of a total divergence.
(B) The total action has the form $S=S_{1}+S_{2}$, $S_{1}=\int\Phi
L_{1}d^{4}x$, $S_{2}=\int\sqrt{-g} L_{2}d^{4}x$ which is the only 
model different from (A) and invariant under SGMF (but now with
$f_{a}=f_{a}(L_{1})$). Similarly, now only $S_{1}$
satisfies
the requirement that the transformation $L_{1}\rightarrow L_{1}+const$
does not effect equations of motion. Both in the case (A) and in the case
(B) it is assumed that $L, L_{1}, L_{2}$ do not depend on $\varphi_{a}$.
(C) The action includes a term which breaks the SGMF symmetry. It is
shown
that in the first order formalism  in cases (A) and (B) the CCP is solved:
 the effective potential vanishes in a
true vacuum state (TVS) without fine tuning. We present a few explicit
field theory models where it is possible to combine the solution of the
CCP with: 1) possibility for inflationary scenario; 2) spontaneously
broken gauge unified theories (including fermions). 
In the case (C),
the
breaking of the SGMF symmetry induces a nonzero energy density for the
TVS.
When considering only
a linear potential for a
scalar field $\phi$ in $S_{1}$, the continuous symmetry
$\phi\rightarrow\phi+const$ is respected. Surprisingly, in this case SSB
takes place while no massless ("Goldstone") boson appears.
We discuss the role of the SGMF symmetry in the possible connection
of this theory with theories of extended objects.
\end{abstract}
 \pagebreak
\section{Introduction}

The cosmological constant problem in the context of general relativity 
(GR) can be explained as follows. In GR one can introduce such a constant 
or one may set it to zero. The problem is that after further 
investigation of elementary particle theory, we discover new phenomena 
like radiative corrections, the existence of condensates, etc. each of 
which contributes to the vacuum energy. In order to have a resulting zero 
or extremely small cosmological constant as required by observations of 
the present day universe , one would have to carefully fine tune 
parameters in the Lagrangian so that all of these contributions more or 
less exactly cancel. This question has captured the attention of many 
authors because, among other things, it could be a serious indication 
that something fundamental has been missed in our standard way of 
thinking about field theory and the way it must couple to gravity. For a 
review of this problem see \cite{CC}.

The situation is made even more serious if one believes in the existence 
of an inflationary phase for the early universe, where the vacuum energy 
plays an essential role. The question is then: what is so special about 
the present vacuum state which was not present in the early universe? In 
this paper we are going to give an answer this question. 

As it is well known, in nongravitational physics the origin from which we
measure energy is not important. For example in nonrelativistic mechanics
a shift in the potential $V\rightarrow V+constant$ does not lead to any
consequence in the equations of motion. In the GR the situation changes 
dramatically. There {\em all} the
energy density, including the origin from which we measure it, affects the
gravitational dynamics.

This is quite apparent when GR is formulated from a variational approach.
There the action is
\begin{equation}
S=\int\sqrt{-g}Ld^{4}x
\label{I1}
\end{equation}
\begin{equation}
L=-\frac{1}{\kappa}R(g)+L_{m}
\label{I2}
\end{equation}
where $\kappa=16\pi G$, $R(g)$ is the Riemannian scalar curvature of the
4-dimensional space-time with metric $g_{\mu\nu}$, $g\equiv
Det(g_{\mu\nu})$ and $L_{m}$ is the matter Lagrangian density.
It is apparent now that the shift of the Lagrangian density $L$,
\quad $L\rightarrow L+C$,\quad $C=const$ is not a symmetry of the action 
(\ref{I1}).
Instead, it leads to an additional piece in the action of the form
$C\int\sqrt{-g}d^{4}x$ which contributes to the equations of motion and
in particular generates a so called "cosmological constant term" in the
equations of the gravitational field. 

In Refs. \cite{GK1}-\cite{GK4} we have developed an approach where the
cosmological constant problem is treated as the absence of gravitational
effects of a possible constant part of the Lagrangian density. The basic
idea is that the measure of integration in the action principle is not
necessarily $\sqrt{-g}$ but it is allowed to "float" and to be determined
dynamically through additional degrees of freedom. In other words the
floating measure is not from first principles related to $g_{\mu\nu}$, 
although relevant equations will in general allow to solve for the new 
measure in terms of other fields of the theory ($g_{\mu\nu}$ and matter 
fields). This theory is based on the demand that such measure respects  
{\em the principle of
non gravitating vacuum energy} (NGVE principle) which states that the
Lagrangian density $L$ can be changed to $L+constant$ without affecting
the dynamics. This requirement is imposed in order to offer a new
approach for the solution of the cosmological constant problem.
Concerning the theories based on the NGVE principle we will refer to
them  as NGVE-theories. 

The invariance $L\longrightarrow L+constant$ for the action is
achieved if the measure of integration in the action is a total
derivative, so that
to an infinitesimal hypercube in 4-dimensional space-time $x_{0}^{\mu}\leq
x^{\mu}\leq x_{0}^{\mu}+dx^{\mu}$, $\mu =0,1,2,3$ we associate a volume
element $dV$ which is: (i) an exact differential, (ii) it is proportional
to
$d^{4}x$ and (iii) $dV$ is a general coordinate invariant. The usual choice,
$\sqrt{-g} d^{4}x$ does not satisfy condition (i).

The conditions (i)-(iii) are satisfied \cite{GK1}, \cite{GK2} if
the measure corresponds to the integration in the space of the four 
scalar fields  $\varphi_{a}, (a=1,2,3,4)$, that is
\begin{equation}
dV =
d\varphi_{1}\wedge
d\varphi_{2}\wedge
d\varphi_{3}\wedge
d\varphi_{4}\equiv\frac{\Phi}{4!}d^{4}x
\label{dV}
\end{equation}
where
\begin{equation}
\Phi \equiv \varepsilon_{a_{1}a_{2}a_{3}a_{4}}
\varepsilon^{\mu\nu\lambda\sigma}
(\partial_{\mu}\varphi_{a_{1}})
(\partial_{\nu}\varphi_{a_{2}})
(\partial_{\lambda}\varphi_{a_{3}})
(\partial_{\sigma}\varphi_{a_{4}}).
\label{Fi}
\end{equation}

Notice that this measure is a particular realization of the NGVE-principle 
(for other possible realization which leads actually to the same results,
see Refs. \cite{GK3}, \cite{GK4}).
For deeper discussion of the geometrical meaning of this realization of the 
measure see Ref.\cite{Hehl} 

We will study three possibilities for the general structure of the action

(A) The most straightforward and complete realization of the NGVE
principle (the so called strong NGVE principle) where the total action is
defined as follows
\begin{equation}
S =\int L\Phi d^{4}x
        \label{Action}
\end{equation}
where $L$ is a total Lagrangian density. We assume in what follows that 
$L$ does not contain explicitly the measure fields, that is the fields
$\varphi_{a}$ by means of which $\Phi$ is defined.

Introducing independent degrees of freedom related to the measure we 
arrive naturally at a conception that all possible degrees of freedom that 
can appear should be considered as such. This is why we expect that the 
first order formalism, where the affine connection is {\em not} assumed to 
be the Christoffel coefficients in general should be preferable to the 
second order formalism where this assumption is made. 

In fact, it is found that the NGVE theory in the context of the first order 
formalism does indeed provide a solution of the cosmological constant 
problem \cite{GK2}, while this is not the case when using the second 
order formalism. 

The simplest example \cite{GK2}-\cite{GK4} (see also Sec.2) where these 
ideas can be tested is that of a matter Lagrangian described by a single 
scalar field with a nontrivial potential. In this case the variational 
principle leads to a constraint which implies the vanishing of the 
effective vacuum energy in any possible allowed configuration of the 
scalar field. These allowed configurations are however constant values at 
the extrema of the scalar field potential and an integration constant 
that results from the equations of motion has the effect of exactly 
canceling the value of the potential at these points. So, the scalar 
field is forced to be a constant and hence  
the theory has no nontrivial dynamics for the scalar field.

In this case the measure (\ref{Fi}) is not determined by the equations of 
motion. In fact a local symmetry (called "Local Einstein Symmetry" (LES)) 
exists which allows us to choose the measure $\Phi$ to be of
whatever we 
want. In 
particular $\Phi=\sqrt{-g}$ can be chosen and in this case the theory 
coincides in the vacuum with GR with $\Lambda =0$.

A richer structure is obtained if a four index field strength which 
derives from a three index potential is allowed in the theory
\cite{GK5},\cite{GK6}. The 
introduction of this term breaks the LES mentioned above. In this case, 
the constraint that the theory provides, allows  to solve for the measure
$\Phi$
in terms of $\sqrt{-g}$ {\em and} the matter fields of the theory. The 
equations can be written in a form that resemble those of the Einstein 
theory by the use of a conformal transformation (or in equivalent language 
by going to the  Einstein frame).

Then the theory which contains a scalar field shows a remarkable feature: 
the effective potential of the scalar field that one obtains in the 
Einstein conformal frame 
is such that  generally allows for an inflationary phase which
evolves at 
a later stage, 
without fine tuning, to a  vacuum of the theory  with zero cosmological 
constant \cite{GK4}. 

The 4-index field strength also allows  for a Maxwell-type dynamics of 
gauge fields and of massive fermions \cite{GK6}.  An explicit construction of
unified gauge theory ($SU(2)\times U(1)$ as an example) based on 
these ideas and keeping all the above mentioned advantages is 
presented in Ref. \cite{GK6}.         

(B) It is possible to check (see \cite{GK1}) that the action 
(\ref{Action})
respects (up to the integral a total divergence) the
infinite dimensional group of shifts of the measure fields $\varphi_{a}$
(SGMF)
\begin{equation}
\varphi_{a}\rightarrow \varphi_{a}+f_{a}(L)
        \label{SG}
\end{equation} 
where $f_{a}(L)$ is an {\em arbitrary} differentiable function of the
total Lagrangian density $L$. Such symmetry in general represents a
nontrivial mixing between the measure fields $\varphi_{a}$ and the matter
and gravitational fields (through L). As it was mentioned in Ref.
\cite{GK1}, this symmetry prevents the appearance of terms of the form
$f(\chi)\Phi$
in the effective action with the single possible exception of $f(\chi)=
c/\chi$ where a scalar $c$ is $\chi$ independent. This
is because in this last  case
the term $f(\chi)\Phi =c\sqrt{-g}$ is $\varphi_{a}$ independent. This
possibility gives rise to the cosmological constant term  in the action 
 while the symmetry (6) is maintained. This
can be generalized to possible contributions of the form
$\int L_{2}\sqrt{-g}d^{4}x$ where $L_{2}$ is $\varphi_{a}$ independent
function of matter fields and gravity if radiative corrections generate a
term $f(\chi)\Phi$ with $f(\chi)=L_{2}/\chi$.

So, let us consider an action \cite{GK7} which
consists of two terms 
\begin{eqnarray}
S & = & S_{1}+S_{2}
\nonumber\\
S_{1} & = & \int L_{1}\Phi d^{4}x
\nonumber\\
S_{2} & = & \int L_{2}\sqrt{-g}d^{4}x
        \label{Action12}
\end{eqnarray}

Now only $S_{1}$ satisfies the requirement that the transformation
$L_{1}\rightarrow L_{1}+const$ does not effect equations of motion, which
is a somewhat weaker version of the NGVE principle. 
In this case, the
symmetry transformation (\ref{SG}) is replaced by 
\begin{equation}
\varphi_{a}\rightarrow \varphi_{a}+f_{a}(L_{1})
        \label{SG1}
\end{equation}

The constraint which appears again in the first order formalism, allows
now to solve
the measure
$\Phi$
in terms of $\sqrt{-g}$ {\em and} the matter fields of the theory
without introducing the four index field strength (see Sec.3). 

In scalar field models with potentials entering in $S_{1}$ and $S_{2}$,
in
the true vacuum state (TVS) $\chi\rightarrow\infty$. However, in the
conformal Einstein frame this singularity does not present and the energy
density of TVS is zero without fine tuning of any  scalar potential in
$S_{1}$ or $S_{2}$ (see Ref.\cite{GK7}). This means that even the weak
version of the NGVE
principle is enough to provide a solution of the CCP. 

We show  in
Subsec.3.2 how it is possible to incorporate gauge fields in such kind of
model.

(C) As it is well known, in order to really understand the role of some
symmetry, one should see what the breaking of such symmetry does. 
With a simple example in Sec.4, we will see that the breaking of the
symmetry
(\ref{SG}) or (\ref{SG1}) can lead to the appearance of a non zero energy
density for the
TVS. In the particular example we study, the additional piece in the
action that is added is of the form
\begin{equation}
S_{3}=-\lambda\int\frac{\Phi^{2}}{\sqrt{-g}}d^{4}x.
        \label{Break}
\end{equation}
which is equivalent to considering of a piece of the Lagrangian
density $L_{1}$ linear in $\chi$. Then
as we show, the TVS energy density appears to be equal to $\lambda$.

In Sec.5 we show that when considering only
a linear potential for a
scalar field $\phi$ in $S_{1}$, the continuous symmetry
$\phi\rightarrow\phi+const$ is respected. Surprisingly, in this case SSB
takes place while no massless ("Goldstone") boson appears. Models with
such a feature exist in each of the cases (A), (B), (C).

\section{The strong NGVE principle}

Starting from the case (A) (see Introduction), we consider  the action
$S=\int\Phi Ld^{4}x$.
Our
choice for the total Lagrangian density is
$L=\kappa^{-1}R(\Gamma,g)+L_{m}$,
where $L_{m}$ is the matter Lagrangian density and $R(\Gamma,g)$ is the
scalar  curvature
$R(\Gamma,g)=g^{\mu\nu}R_{\mu\nu}(\Gamma)$ of the space-time of the
affine connection  $\Gamma^{\mu}_{\alpha\beta}$: \,
$R_{\mu\nu}(\Gamma)=R^{\alpha}_{\mu\nu\alpha}(\Gamma)$, \,
$R^{\lambda}_{\mu\nu\sigma}(\Gamma)\equiv 
\Gamma^{\lambda}_{\mu\nu,\sigma}-\Gamma^{\lambda}_{\mu\sigma,\nu}+
\Gamma^{\lambda}_{\alpha\sigma}\Gamma^{\alpha}_{\mu\nu}-
\Gamma^{\lambda}_{\alpha\nu}\Gamma^{\alpha}_{\mu\sigma}$.
This curvature tensor is invariant under the
$\lambda$- transformation \cite{EK}
$\Gamma^{\prime 
\mu}_{\alpha\beta}=\Gamma^{\mu}_{\alpha\beta}+
\delta^{\mu}_{\alpha}\lambda,_{\beta}$. \,
In the
NGVE-theory, it  allows us to eliminate the contribution to the
torsion which appears as a result of introduction of the new measure.
However, even after this
still 
there is the non metric contribution to the 
connection related to the measure (it is expressed in terms of 
derivatives the scalar field $\chi\equiv\Phi/\sqrt{-g}$).

In addition to this, in the vacuum and in some matter models, the theory
possesses a local symmetry which plays a major role. This symmetry
consists of a conformal transformation of the metric
$g_{\mu\nu}(x)=J^{-1}(x)g^{\prime}_{\mu\nu}(x)$ accompanied by a 
corresponding
diffeomorphism $\varphi_{a}\longrightarrow\varphi^{\prime}_{a}=
\varphi^{\prime}_{a}(\varphi_{b})$ in the space of the scalar fields
$\varphi_{a}$ such that  $J=
Det(\frac{\partial\varphi^{\prime}_{a}}{\partial\varphi_{b}})$.
Then for $\Phi$ we have: $ \Phi(x)=J^{-1}(x)\Phi^{\prime}(x)$. In the
presence of fermions this symmetry is appropriately generalized \cite{GK2}.
For models where it
holds, it is possible to choose the gauge where the measure $\Phi$
coincides with $\sqrt{-g}$, the measure of GR. This is
why we call this symmetry {\em "local Einstein symmetry"} (LES).

 Varying the action with respect to $\varphi_{a}$ we get
$A^{\mu}_{b}\partial_{\mu}\lbrack -\frac{1}{\kappa}R(\Gamma,g)+
L_{m}\rbrack =0$ where $A^{\mu}_{b}=\varepsilon_{acdb}
\varepsilon^{\alpha\beta\gamma\mu}
(\partial_{\alpha}\varphi_{a})
(\partial_{\beta}\varphi_{c})
(\partial_{\gamma}\varphi_{d})$.
If $Det (A_{b}^{\mu}) =
\frac{4^{-4}}{4!}\Phi^{3}\neq 0$ then
\begin{equation}
-\frac{1}{\kappa}R(\Gamma,g)+L_{m}=M=const
\label{1}
\end{equation}

Performing the variation with respect to
$g^{\mu\nu}$ we get (here for simplicity we don't consider fermions)
 \begin{equation}
-\frac{1}{\kappa}R_{\mu\nu}(\Gamma)+\frac{\partial L}{\partial g^{\mu\nu}}=0
\label{2}
\end{equation}

Contracting eq.(\ref{2}) with $g^{\mu\nu}$ and making use eq.(\ref{1}) we
get the constraint
\begin{equation}
g^{\mu\nu}\frac{\partial(L_{m}-M)}{\partial g^{\mu\nu}}-(L_{m}-M)=0
\label{3}
\end{equation}

For the cases where the LES is an exact symmetry, we
can eliminate the mentioned above $\chi$-contribution to the connection.
Indeed, for  $J=\chi$ we get $\chi^{\prime}\equiv 1$ and
$\Gamma^{\prime \alpha}_{\mu\nu}=
\{ ^{\alpha}_{\mu\nu}\}^{\prime}$, where
$\{ ^{\alpha}_{\mu\nu}\}^{\prime}$
 are the Christoffel's coefficients corresponding
to the new metric $g^{\prime}_{\mu\nu}$. In this gauge the affine space-time
becomes a Riemannian space-time.

When applying the theory to the matter model of a single scalar field with a 
potential $V(\varphi)$, the constraint (\ref{3}) implies $V(\varphi)+M=0$, 
which means that $\varphi$ is a constant (the equation of motion of 
$\varphi$ implies also that $V^{\prime}(\varphi)=0$). Since 
$\varphi= constant$, Eq. (\ref{2}) implies $R_{\mu\nu}(\Gamma,g)=0$ and 
also Eq. (\ref{1}) implies $R(\Gamma,g)=0$, if $V(\phi)+M=0$ is taken 
into account. $V(\phi)+M=0$ dictates the vanishing of the LES violating 
terms and disappearance of dynamics of $\varphi$, so LES is effectively 
restored on the mass shell. Choosing the gauge $\chi =1$ we see that the 
potential does not have a gravitational effect since the standard Ricci 
tensor vanishes and flat space-time remains the only natural vacuum.

The above solution of the CCP is at the price 
of the elimination of a possible scalar field dynamics. We have shown in
\cite{GK5},\cite{GK6} 
that the introduction of a 4-index field strength can restore normal 
scalar, gauge and fermion dynamics and as a bonus provide:
 (i) {\em the possibility of inflation in the early 
universe with a transition (after reheating) to a $\Lambda =0$ phase without 
fine tuning} and
(ii) {\em a solution to the hierarchy problem in the context of unified 
gauge dynamics with SSB}.

A four index field strength is derived from a three index gauge 
potential according to 
$F_{\mu\nu\alpha\beta}=\partial_{[\mu}A_{\nu\alpha\beta]}$.
The physical scenario we have in mind is one where all gauge fields, 
including $A_{\nu\alpha\beta}$ are treated in a unified way. The possible 
physical origin of $A_{\nu\alpha\beta}$ can be for example an effective 
way to describe the condensation of a vector gauge field in some extra 
dimensions \cite{G}. In this picture for example this means that all 
gauge fields should appear in a combination having all the same 
homogeneity properties with respect to conformal transformations of the 
metric. This is achieved if all dependence on field strengths is through 
the "gauge fields complex" \,
$y=F^{a}_{\mu\nu}F^{a\mu\nu}+
\frac{\varepsilon^{\mu\nu\alpha\beta}}{\sqrt{-g}}\partial_{\mu}
A_{\nu\alpha\beta}$.

Considering for illustration only one vector gauge field $\tilde{A}_{\mu}$ 
and 
a charged scalar field $\phi$, we take a generic action satisfying the 
above requirements in the unitary gauge ($\phi =\phi^{*};\, |\phi|=
\frac{1}{\sqrt{2}}\varphi$)
\begin{equation}
S=\int\Phi d^{4}x\left[-\frac{1}{\kappa}R(\Gamma ,g)-m^{4}f(u)
+\frac{1}{2}g^{\mu\nu}\partial_{\mu}\varphi\partial_{\nu}\varphi
-V(\varphi)+\frac{1}{2}\tilde{e}^{2}\varphi^{2}g^{\mu\nu}\tilde{A}_{\mu}
\tilde{A}_{\nu}\right],
\label{5}
\end{equation} 
where $u\equiv y/m^{4}$,\, $m$ is a mass parameter 
and $f(u)$ is a nonspecified function which has to have an extremum at 
some point $u=u_{0}>0$ to provide physically reasonable consequences (see 
below).  This is motivated from the well known instability of nonabelian
gauge theories as found by Savvidy \cite{Sav} and which leads naturally to
such type of functions.

The equations of motion obtained from the $A_{\nu\alpha\beta}$ variation
imply
$\chi f^{\prime}=\omega =constant$
where $\omega$ is a dimensionless integration constant.
The constraint (\ref{3}) becomes now
\begin{equation}
-2uf^{\prime}(u)+f(u)+\frac{1}{m^{4}}[V(\varphi)+M]=0,
\label{7}
\end{equation}
which allows to find $u=u(\varphi)$.

One can then see that all equations can be put in the standard GR, scalar 
field and gauge field form if we make the conformal transformation to the 
"Einstein frame"
$\overline{g}_{\mu\nu}=\chi g_{\mu\nu}$.
In the Einstein frame, the scalar field acquires an effective potential 
\begin{equation}
V_{eff}(\varphi)=\frac{y}{\omega}\left(f^{\prime}(u)\right)^{2}
\label{9}
\end{equation}

If there is a point $u=u_{0}$ where 
$f^{\prime}(u_{0})=0$, then if $y_{0}/\omega >0$, such a state is a 
stable vacuum of the theory. This vacuum state is defined by the gauge 
and scalar condensates 
$(u_{0}, \varphi_{0})$ connected 
by the relation $f(u_{0})+\frac{1}{m^{4}}[V(\varphi_{0})+M]=0$,
representing the exact cancellation of the contributions to the vacuum 
energy of the scalar field and of gauge field condensate. Therefore the 
effective cosmological constant in this vacuum becomes zero without fine 
tuning.

One can see also that $\frac{dV_{eff}}{d\varphi}=\frac{1}{\omega}
\frac{df}{du}\frac{dV}{d\varphi}$, so another extremum where 
$V^{\prime}=0$ for example can serve as a phase with nonzero effective 
cosmological constant and therefore inflation becomes possible as well.
This vacuum 
is smoothly connected by dynamical evolution of the scalar field, with 
the zero cosmological constant phase, thus providing a way to achieve 
inflation and transition (after standard reheating period) to $\Lambda =0$
phase without fine tuning.

Stability of gauge fields requires $\omega >0$. In this case theory 
acquires canonical form if the new fields and couplings are defined
$A_{\mu}=2\sqrt{\omega}\tilde{A}_{\mu}$, \, 
$e=\frac{\tilde{e}}{2\sqrt{\omega}}$. Appearance of the VEV of the scalar 
field $\varphi_{0}$ leads to 
the standard Higgs mechanism.

Fermions can also be introduced \cite{GK6} in such a way that normal 
massive propagation is obtained in the Einstein frame. The resulting fermion 
mass 
is proportional to $\frac{\varphi_{0}}{\sqrt{u_{0}\omega}}$. The gauge boson 
mass, depending on $e$ also goes as $\propto{1}/{\sqrt{\omega}}$. So we 
see that a big value of the integration constant $\omega$ pushes both 
effective masses and gauge coupling constants to small values, thus 
providing a new approach to the solution of the hierarchy problem. 
Furthermore we see that fermion masses include additional factor 
$u^{-1/2}_{0}$. Therefore, if the gauge complex condensate $u_{0}$ is 
big enough, it can explain why fermion masses are much less then boson 
ones. 

Finally, there is no obstacle for the construction of a realistic unified 
theories (like electroweak, QCD, GUT) along the lines of the simple example 
displayed above \cite{GK6}. The common feature of such theories is the
fact 
that the 
stable vacuum developed after SSB has zero effective cosmological constant.

\section{Models satisfying the weak NGVE principle}
\subsection{The simplest model}

To demonstrate how the theory works when it is based on the weak NGVE
principle (see the case(B) in Introduction), we start here from the
simplest 
model \cite{GK7} in the first order formalism including scalar field
$\phi$ and gravity
according to the prescription of the NGVE principle and in addition to 
this we include the standard cosmological constant term. 
So, we consider an action 
\begin{equation}
S=\int L_{1}\Phi d^{4}x + \int\Lambda\sqrt{-g}d^{4}x
\label{AW}
\end{equation}
where 
$L_{1}=-\frac{1}{\kappa}R(\Gamma,g)+
\frac{1}{2}g^{\mu\nu}\phi_{,\mu}\phi_{,\nu}
-V(\phi)$.

Performing the variation with respect to the measure fields $\varphi_{a}$ 
 we obtain equations 
$A^{\mu}_{a}\partial_{\mu}L_{1}=0$
where $A^{\mu}_{b}=\varepsilon_{acdb}
\varepsilon^{\alpha\beta\gamma\mu}
(\partial_{\alpha}\varphi_{a})
(\partial_{\beta}\varphi_{c})
(\partial_{\gamma}\varphi_{d})$.
If $\Phi\neq 0$, 
then it follows from the last equations that $L_{1}=M=const$

Varying the action (\ref{AW}) with respect to $g^{\mu\nu}$ we get
\begin{equation}
\Phi(-\frac{1}{\kappa}R_{\mu\nu}(\Gamma)+\frac{1}{2}\phi_{,\mu}\phi_{,\nu})
-\frac{1}{2}\sqrt{-g}\Lambda g_{\mu\nu}=0
\label{GW}
\end{equation}

Contracting Eq.(\ref{GW}) with $g^{\mu\nu}$ and using equation $L_{1}=M$
we 
obtain the constraint
\begin{equation}
M+V(\phi)-\frac{2\Lambda}{\chi}=0
\label{CW}
\end{equation}
where again $\chi\equiv\Phi/\sqrt{-g}$.

The scalar field $\phi$ equation is
$(-g)^{-1/2}\partial_{\mu}(\sqrt{-g}g^{\mu\nu}\partial_{\nu}\phi)+
\sigma_{,\alpha}\phi^{,\alpha}+V^{\prime}=0$
where $\sigma\equiv\ln\chi$ and $V^{\prime}\equiv dV/d\phi$.

The derivatives of the field $\sigma$ enter both the gravitational 
equation (\ref{GW}) (through the connection) and in the scalar field
equation.
 By a conformal transformation 
$g_{\mu\nu}\rightarrow\overline{g}_{\mu\nu}=\chi g_{\mu\nu}; \quad 
\phi\rightarrow\phi$
to an "Einstein picture" and using the constraint (\ref{CW}) we obtain the 
canonical form of equations for the scalar field \\ 
$(-\overline{g})^{-1/2}\partial_{\mu}(\sqrt{-\overline{g}}\quad
\overline{g}^{\mu\nu}\partial_{\nu}\phi)+V_{eff}^{\prime}(\phi)=0$
and the gravitational equations in the Riemannian space-time with metric 
$\overline{g}_{\alpha\beta}$:\quad
$R_{\mu\nu}(\overline{g}_{\alpha\beta})-
\frac{1}{2}\overline{g}_{\mu\nu}R(\overline{g}_{\alpha\beta})
=\frac{\kappa}{2}T^{eff}_{\mu\nu}(\phi)$
where $T^{eff}_{\mu\nu}(\phi)=\phi_{,\mu}\phi_{,\nu}-
\frac{1}{2}\overline{g}_{\mu\nu}
\phi_{,\alpha}\phi_{,\beta}\overline{g}^{\alpha\beta}+
\overline{g}_{\mu\nu}V_{eff}(\phi)$, and
$V_{eff}(\phi)= 
\frac{1}{4\Lambda}[M+V(\phi)]^{2}$

We see that for any analytic function $V(\phi)$, the effective potential 
in the Einstein picture has an extremum, i.e. $V_{eff}^{\prime}=0$, 
either when $V^{\prime}=0$ or $V+M=0$. The extremum $\phi=\phi_{1}$ where 
$V^{\prime}(\phi_{1})=0$ has nonzero energy density 
$[M+V(\phi_{1})]^{2}/4\Lambda$ if a fine tuning is not assumed. In 
contrast to this, if $\Lambda >0$, the state $\phi =\phi_{0}$ where 
$V(\phi_{0})+M=0$ is the absolute minimum and therefore $\phi_{0}$ is a 
true vacuum with zero cosmological constant without any fine tuning. A mass 
square of the scalar field describing small fluctuations around 
$\phi_{0}$ is
$m^{2}=\frac{1}{2\Lambda}[V^{\prime}(\phi_{0})]^{2}$.
Exploiting the possibility to choose any analytic $V(\phi)$, we can pick 
the structure of $V_{eff}(\phi)$ so that it
allows for an
inflationary era, the possibility of reheating after scalar field
oscillations and the setting down to a zero cosmological constant phase
at the later stages of cosmological evolution, without fine tuning.

Notice that if $V+M$ achieves the value zero at some value of $\phi
=\phi_{0}$, this point represents the absolute minimum of the effective
potential. If this is not the case for a particular choice of potential
$V$
and of integration constant $M$, it is
always possible to choose an infinite range of values of $M$ where this
will happen. Therefore no fine tuning of parameters has to be invoked,
the zero value of the true vacuum energy density appears naturally in this
theory. 

Further generalizations, like considering a term of the form 
$\int U(\phi)\sqrt{-g}d^{4}x$ instead of 
$\Lambda\int \sqrt{-g}d^{4}x$, the possibility of coupling of scalar 
fields to curvature, etc. do not modify the qualitative nature of the 
effects described here and they will be studied in a more detailed 
publication. For example, even in the presence of a generic 
$U(\phi)$ the resulting effective potential vanishes when 
$V(\phi_{0})+M=0$ and it goes as 
$V_{eff}=\frac{1}{4U(\phi_{0})}(V+M)^{2}$ in the region $V+M\sim 0$.

\subsection{Including the gauge fields}

The weak NGVE principle allows to incorporate  gauge fields in a way
which is  simpler  than in the context of the strong NGVE
principle. Taking in Eqs. (\ref{Action12})
\begin{eqnarray}
L_{1} & = & -\frac{1}{\kappa}R(\Gamma, g)+g^{\mu\nu}(\partial_{\mu} 
-ieA_{\mu})\phi (\partial_{\nu}+ieA_{\nu})\phi^{*}-V(\phi)
\nonumber\\
L_{2} & = &
-\frac{1}{4}g^{\alpha\beta}g^{\mu\nu}F_{\alpha\mu}F_{\beta\nu}+\Lambda
\label{LW}
\end{eqnarray}
we see that except of the $V$-term in $L_{1}$ and $\Lambda$-term in 
$L_{2}$, the action (\ref{Action12}) is invariant under the LES (notice
that in $S_{2}$ the LES 
takes the form of the conformal transformation  of the metric
$g_{\mu\nu}(x)=J^{-1}(x)g^{\prime}_{\mu\nu}(x)$). Therefore, as one can
check explicitly, the gauge
field does not enter in the constraint which turns out to be identical to  
 Eq. (\ref{CW}). 

For the gauge field in the unitary gauge we get 
$(-g)^{-1/2}\partial_{\mu}(\sqrt{-g}g^{\mu\alpha}g^{\nu\beta}F_{\alpha\beta})
+ \\ e^{2}\chi\varphi^{2}g^{\mu\nu}A_{\mu}=0$. In the Einstein
conformal frame where the metric is $\overline{g}_{\mu\nu}=\chi
g_{\mu\nu}$, the gauge field equation takes the canonical form
\begin{equation}
\frac{1}{\sqrt{-\overline{g}}}
\partial_{\mu}(\sqrt{-\overline{g}}\enspace\overline{g}^{\mu\alpha}
\overline{g}^{\nu\beta}F_{\alpha\beta})+e^{2}\varphi^{2}
\overline{g}^{\mu\nu}A_{\mu}=0
\label{GE}
\end{equation}

The gravitational and scalar field equations including the effective
scalar field potential in the Einstein picture
coincide with  the appropriate equations of the model in the previous
subsection 3.1.  Keeping the assumption that 
$\Lambda >0$ we can see that all conclusions
concerning the vanishing of the
vacuum energy in the true vacuum state $\varphi =\varphi_{0}$ where
$V(\varphi_{0})+M=0$ remain unchanged. In addition to this, the Higgs
mechanism for the mass generation in the true vacuum state $\varphi_{0}$
works now in a regular way in contrast to the model based on the strong
NGVE principle where the effective coupling constant (and as a
consequence, the mass of the gauge boson) depends on the integration
constant.  

\section{The true vacuum energy density
as an effect of the SGMF symmetry breaking}

In the previous section we have seen that the presence of the cosmological
constant term in the original action of the weak NGVE theory does not
change the result of the strong NGVE theory: the vacuum energy density of
the TVS (in the Einstein picture) is zero. We are going to show now that
the appearance of the nonzero vacuum energy density in the TVS (that
is the effective cosmological term) in the Einstein picture  can be the
result of the explicit breaking of the SGMF symmetry (\ref{SG}) or
(\ref{SG1}) by adding the simplest form of a SGMF symmetry breaking term
(\ref{Break}) to the action (\ref{Action12}). Since such term is invariant
under the LES, it does not contribute to the constraint as one can check
explicitly, and therefore the constraint coincides with Eq. (\ref{CW}).

In this case the gravitational equations in the Einstein frame which still
is defined by $\overline{g}_{\mu\nu}=\chi g_{\mu\nu}$ are
\begin{equation}
R_{\mu\nu}(\overline{g})-\frac{1}{2}\overline{g}_{\mu\nu}R(\overline{g})=
\frac{\kappa}{2}\lbrace\phi,_{\mu}\phi,_{\nu}-
\frac{1}{2}\overline{g}_{\mu\nu}\phi,_{\alpha}\phi,_{\beta}
\overline{g}^{\alpha\beta}+\lbrack\frac{1}{4\Lambda}(V+M)^{2}
+\lambda\rbrack\overline{g}_{\mu\nu}\rbrace
\label{GB}
\end{equation}

In the TVS
$\phi=\phi_{0}$ 
where
$V(\phi_{0})+M=0$ 
the last term
in Eq.
(\ref{GB}) acts 
as an effective 
cosmological
term.

\section {Model with continuous symmetry related to the NGVE principle
and SSB
without generating a massless scalar field}

It is interesting to see what happens in the above models for the choice
$V=J\phi$, where $J$ is some constant. For simplicity we consider here 
the case (B) (see Sec. 3.1). Then the action (\ref{AW}) is
invariant (up to the integral of total divergence) under the shift
$\phi\rightarrow\phi +const$ which is in fact the symmetry
$V\rightarrow V+const$ related to the NGVE principle. Notice that if we
 consider
the model with complex scalar field $\psi$, where $\phi$ is the phase of
$\psi$, then the symmetry $\phi\rightarrow\phi +const$ would be the
$U(1)$ - symmetry.

The effective potential $V_{eff}(\phi)=
\frac{1}{4\Lambda}[M+V(\phi)]^{2}$ in such a model has the form
\begin{equation}
V_{eff}=\frac{1}{2}m^{2}(\phi -\phi_{0})^{2}
\label{18}
\end{equation}
 where $\phi_{0}=-M/J$
and $m^{2}=J^{2}/2\Lambda$. We see that the symmetry
$\phi\rightarrow\phi +const$ is spontaneously broken and mass
generation
is obtained. However, {\em no massless scalar field results from
the
process of SSB in this case, i.e. Goldstone theorem does not apply
here}.

This seems to be a special feature of the NGVE - theory which
allows:
1) To start with linear potential $J\phi$ without destroying the
shift
symmetry $\phi\rightarrow\phi +const$, present in the
$\partial_{\mu}\phi \partial^{\mu}\phi$ piece, due to the coupling to the
dynamical measure (\ref{Fi}). This shift symmetry is now a symmetry of
the
action up to a total divergence.
2) This potential gives rise to an effective potential
$(M+J\phi)^{2}/4\Lambda$. The constant of integration $M$ being
responsible for the SSB.

In the case (C) (Sec.4) for the choice $V=J\phi$, the only difference is
the appearance of the $\lambda$-term in addition to Eq.(\ref{18}).
Similar effect can be obtained also in the strong NGVE - theory 
(without introduction of an explicit $\Lambda$-term, as in Sec.2) but
with the
use of 4-index field strength condensate. The possibility of constructing
spontaneously broken $U(1)$ models which do not lead to associated
Goldstone bosons is of course of significant physical relevance. One may
recall for example the famous $U(1)$ problem in QCD \cite{U1}. Also
the possibility of mass generation for axions is of considerable
interest. These
issues will be developed further in elsewhere \cite{prep}.

\section{Discussion and Conclusions}

We have seen that the consideration of a measure $\Phi$ independent of
$\sqrt{-g}$, which in fact means that its relation to $g_{\mu\nu}$ and
other fields is to be found dynamically, provides a new  approach to the
resolution of the CCP. In fact the consideration of this dynamical
measure, in the context of the first order formalism solves the CCP in a
wide range of models where the structures $S_{1}$ and $S_{2}$ of
Eq.(\ref{Action12}) are allowed  to appear, thus ensuring that the SGMF
symmetry (\ref{SG}) or (\ref{SG1}) are satisfied.

The violation of the SGMF symmetry leads in contrast to an explicit
appearance of an energy density in the TVS as we have seen in Sec.4 in a
specific example. If this violation appears as a result of quantum
corrections, it represents then the appearance of an anomaly of the  SGMF
symmetry (\ref{SG}) or (\ref{SG1}). We then
have a reason for the smallness of such terms if not of their absolutte
vanishing ( in the case of exact symmetry). This resembles the situation
of quark masses and chiral invariance (CI). In this case, as it is well
known, CI forbids a quark mass. If a quark mass nevertheless appears, CI
ensures that quark masses remain small even after the consideration of
quantum corrections. In a similar way if a small SGMF symmetry
breaking term appears, hopefully, it will not be renormalized into a large
contribution after quantum effects are considered.

Since the main results of this work seem to depend on the existence of the
SGMF symmetry, it is interesting to see the possible interpretations of
this symmetry.

First of all notice that when $\Phi\neq 0$, $L_{1}=const.$ and the
symmetry (\ref{SG1}) represents just a trivial shift of $\varphi_{a}$ by a
constant. This shift becomes nontrivial if we allow for points where
$\Phi =0$ and $L_{1}$ can vary in such "defect" points. Such defects can
be naturally associated with extended objects and the nontrivial symmetry
(\ref{SG1}) appears then like invariance under reparametrization of the
extended object coordinates $\varphi_{a}$.

A complete group-theoretical and/or algebraic study of the infinite
dimensional symmetries of the theory has not been carried out yet and
this should be an interesting exercise. One should notice that in
addition to (\ref{SG1}) there is the infinite dimensional symmetry of
volume preserving internal diffeomorphisms (VPD)
$\varphi^{\prime}_{a}=\varphi^{\prime}_{a}(\varphi_{b})$ such that
$\Phi^{\prime}=\Phi$. Such transformations applied after a transformation
of the form (\ref{SG1}) lead to something new. The full symmetry group of
transformation contains element that are not in SGMF and not in VPD which
are only subgroups of the yet unknown full group of internal symmetries of
the measure scalars $\varphi_{a}$.

In our treatment of the symmetry $L_{1}\rightarrow L_{1}+const.$ we have
ignored possible topological effects, since $\int\Phi d^{4}x$ being a
total divergence can be responsible however for topological effects,
similar to the well known $\Theta$-term in QCD.

As we have seen in Sec.5, such symmetry can be exploited to
construct (through the introduction of a linear potential $V=J\phi$
coupled to the measure $\Phi$) a theory that is globally $U(1)$-invariant
with SSB and yet without appearance of a Goldstone boson. Possible
applications to axion mass generation, etc has to be explored.

Finally, one can see that in the presence of matter which does not satisfy
automatically the constraint (\ref{3}), that is it is not LES-invariant,
such matter will influence the effective potential of a scalar field in
the Einstein picture, through the constraint. For example, including dust
in $L_{1}$ of the model of Sec.3.1, it is found that the constraint links
$\Lambda$ to the amount of dust in the homogeneous cosmological solutions
\cite{prep}  

\section{Acknowledgments}
We thank J. Bekenstein, R. Brustein, A. Davidson, F. Hehl, Y. Ne'eman
and J. Portnoy for  interesting discussions on the subjects of this paper.
We also thank the organizers of the fourth Alexander Friedmann
International Seminar on Gravitation and Cosmology Professors Yu. N.
Gnedin, A. A. Grib and V. M. Mostepanenko for providing a stimulating
atmosphere during this conference.

\end{document}